\documentclass{natcomSI}
\usepackage{graphicx}
\usepackage{bm}
\usepackage{amssymb}
\usepackage{microtype}
\usepackage{xfrac}
\usepackage{makecell}
\usepackage{array}
\usepackage[version=4]{mhchem}
\usepackage{gensymb}
\usepackage[colorlinks=true]{hyperref}

\usepackage{comment}

\newcommand{\angstrom}{\text{\normalfont\AA}}

\title{Incommensurate magnetism mediated by Weyl fermions in NdAlSi}

\author{Jonathan~Gaudet${}^{1}$,
Hung-Yu~Yang${}^{2}$,
Santu~Baidya${}^{3}$,
Baozhu~Lu${}^{4}$,
Guangyong~Xu${}^{5}$,
Yang~Zhao${}^{6,5}$,
Jose~A.~Rodriguez${}^{6,5}$,
Christina~M.~Hoffmann${}^{7}$,
David~E.~Graf${}^{8}$,
Darius~H.~Torchinsky${}^{4}$,
Predrag~Nikolić${}^{9,1}$,
David~Vanderbilt${}^{3}$,
Fazel~Tafti${}^{2}$,
Collin~L.~Broholm${}^{1,5}$
}
\begin{document}

\maketitle

\begin{affiliations}
 \item Department of Physics and Astronomy and Institute for Quantum Matter, The Johns Hopkins University, Baltimore, Maryland 21218, USA
 \item Department of Physics, Boston College, Chestnut Hill, MA 02467, USA
 \item Department of Physics and Astronomy, Rutgers University, Piscataway, New Jersey 08854, USA
 \item Department of Physics, Temple University, Philadelphia, PA 19122, USA
 \item NIST Center for Neutron Research, 100 Bureau Drive, National Institute of Standards and Technology, Gaithersburg, MD 20899-6102, USA
 \item Department of Materials Science and Engineering, University of Maryland, College Park, Maryland 20742, USA
 \item Neutron Scattering Division, Oak Ridge National Laboratory, Oak Ridge, Tennessee 37831, USA
 \item National High Magnetic Field Laboratory, Tallahassee, FL 32310, USA
 \item Department of Physics and Astronomy, George Mason University, fairfax, VA 22030, USA

\end{affiliations}

\begin{abstract}
Emergent relativistic quasiparticles in Weyl semimetals are the source of exotic electronic properties such as surface Fermi arcs, the anomalous Hall effect, and negative magnetoresistance, all observed in real materials. Whereas these phenomena highlight the effect of Weyl fermions on the electronic transport properties, less is known about what collective phenomena they may support. Here, we report a new Weyl semimetal, NdAlSi that offers an example. Using neutron diffraction, we report a long-wavelength magnetic order in NdAlSi whose periodicity is linked to the nesting vector between two topologically non-trivial Fermi pockets, which we characterize using density functional theory and quantum oscillation measurements. Our work provides a rare example of Weyl fermions driving collective magnetism.
\end{abstract}

\newpage

\section{Introduction}

Weyl semimetals are topologically nontrivial phases of matter that sustain low energy excitations in the form of massless fermionic quasiparticles known as the Weyl fermions~\cite{Armitage2018,Wan2015}. It is necessary to break either inversion or time-reversal symmetry to establish a Weyl semimetal~\cite{Lv2015,Yang2015,Xu2015,Kim2018,Liu2019,Liu2019b,Belopolski2019}. A rare occasion arises, however, if a material breaks both symmetries and offers an opportunity to study the interplay between magnetism and Weyl fermions~\cite{Chang2018,Suzuki2019,Puphal2020,Yang2020}. Here we present an extensive experimental and theoretical study of a new Weyl semimetal, NdAlSi, that breaks both symmetries. Our neutron diffraction experiment reveals the leading instability of the Weyl semimetal is to long-wavelength incommensurate order followed by a lower temperature transition to commensurate ferrimagnetism. Using quantum oscillation measurements and density functional theory, we find that the incommensurate wavevector connects different branches of the Fermi surface that contain Weyl fermions. This raises the interesting possibility of a magnetic order driven by relativistic electrons in NdAlSi.

\begin{figure}[t]
    \includegraphics[width=400pt]{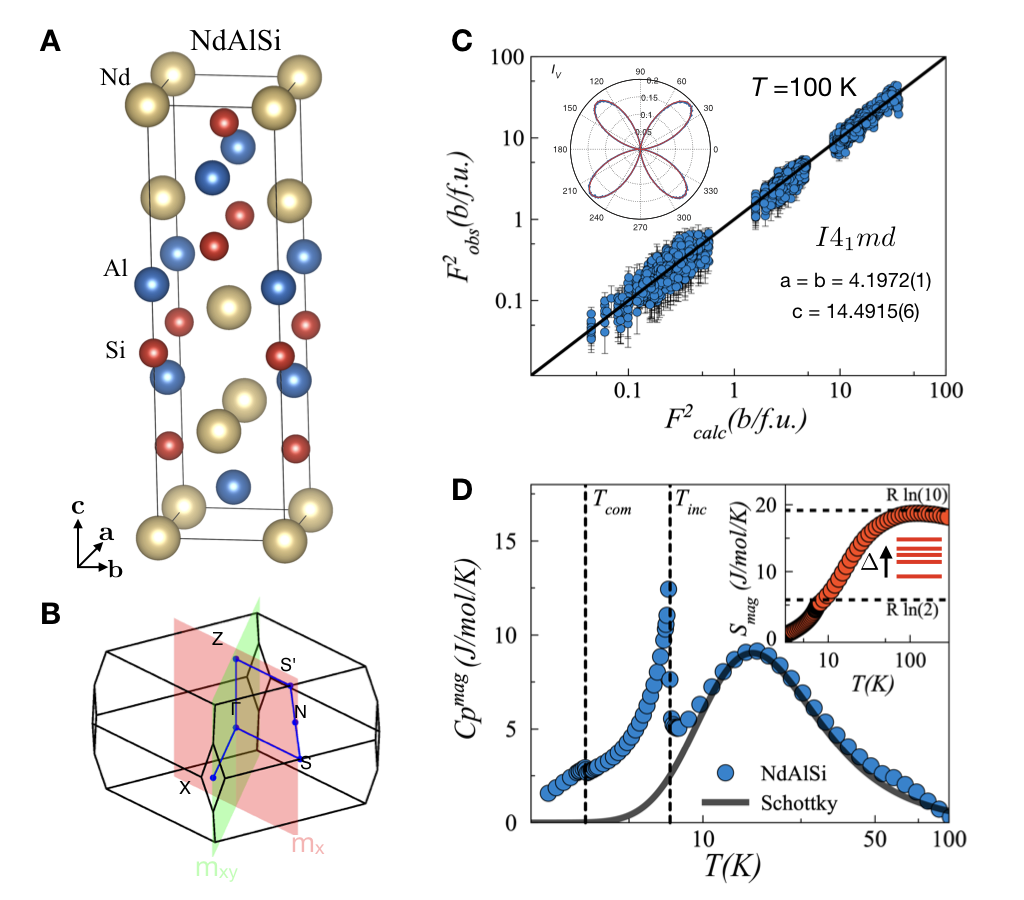}
    \centering
    \caption{\textbf{Crystal strucure of NdAlSi.}(A) Structure of the non-centrosymmetric NdAlSi with space group $I4_1md$ (109) and its associated first Brillouin zone shown in (B). There are two simple mirror planes, $m_{x}$ and $m_{y}$, and two glide mirrors, $m_{xy}$ and $m_{x\bar{y}}$. These are shown in (B) as the light red and green planes, with the  $Z$--$\Gamma$--$X$ path lying in the green $m_{xy}$ mirror plane and the $\Gamma$--$S$--$N$--$S^{\prime}$--$Z$--$\Gamma$ path lying in the red $m_{x}$ mirror plane. (C) Refinement of single crystal nuclear neutron diffraction data collected at 100~K. The Second Harmonic Generation signal is shown as an inset. (D) Temperature dependence of the magnetic heat capacity and magnetic entropy of NdAlSi, which were isolated from the net  heat capacity by subtracting the heat capacity of non-magnetic LaAlSi.}
    \label{fig:Bulk1}
\end{figure}

\section{Structural, Magnetic and Electronic Bulk Properties}

Single crystals of NdAlSi were grown by a self-flux approach (see methods). The body-centered tetragonal unit cell and the first Brillouin zone (BZ) of NdAlSi are shown in Fig.~\ref{fig:Bulk1}A and B. The structure has two vertical mirror planes ($m_{x}$ and $m_{y}$) but lacks a horizontal mirror plane ($m_{z}$), and thus breaks inversion symmetry (Fig.~\ref{fig:Bulk1}B). Site mixing between Al and Si can, however, restore the $m_{z}$ mirror plane and change the space (point) group from non-centrosymmetic $I4_1md$ ($C_{4v}$) to centrosymmetric $I4_1/amd$ ($C_{4h}$). Contrary to x-ray scattering, the neutron scattering length of Al and Si are sufficiently different ($b({\rm Al})=3.449~{\rm fm}, b({\rm Si})=4.1491~{\rm fm})$ that site mixing is readily apparent in neutron diffraction. The refinement of our single crystal neutron diffraction pattern for NdAlSi in the $I4_1md$ space group is shown in Fig.~\ref{fig:Bulk1}C. A better fit  ($\chi^2=5.70$) is obtained in $I4_1md$ compared to refinement in the $I4_1/amd$ space group ($\chi^2~=~6.04$), which yields a limit of 9\% on Si-Al site mixing. The inset of Fig.~\ref{fig:Bulk1}C shows a strong second harmonic generation (SHG) signal ($\chi_{xxz}$ = -115(3)~pm/V, $\chi_{zxx}$ = 94(2)~pm/V and $\chi_{zzz}$ = 564(5)~pm/V) that originates from a bulk electric dipole and fits to the point group $C_{4v}$. Thus, we confirm the non-centrosymmetric space group $I4_1md$ as the correct structure for NdAlSi (Fig.S1).

The magnetic heat capacity ($C^{mag}_p$) of NdAlSi is plotted in Fig.~\ref{fig:Bulk1}D, revealing a broad anomaly at approximately 18~K, as well as two peaks at $T_{\mathrm{inc}}=7.2(1)$~K and $T_{\mathrm{com}}=3.3(1)$~K. A total entropy of $\Delta S=0.96(2) \times R\ln(10)$ is released between 2.35~K and 300~K (inset of Fig.~\ref{fig:Bulk1}D), which is the value expected for the ground-state spin-orbital manifold of Nd$^{3+}$ with $L=6$, $S=3/2$, and $J=9/2$. Fitting $C^{mag}_p(T)$ indicates splitting of the $2J+1=10$-fold degenerate $J$-multiplet into a ground state doublet and four excited doublets at $\Delta=4(2)$~meV. The low temperature anomalies at $T_{\mathrm{inc}}=7.2$~K and $T_{\mathrm{com}}=3.3$~K mark two magnetic phase transitions.

\begin{figure}[t]
    \includegraphics[width=420pt]{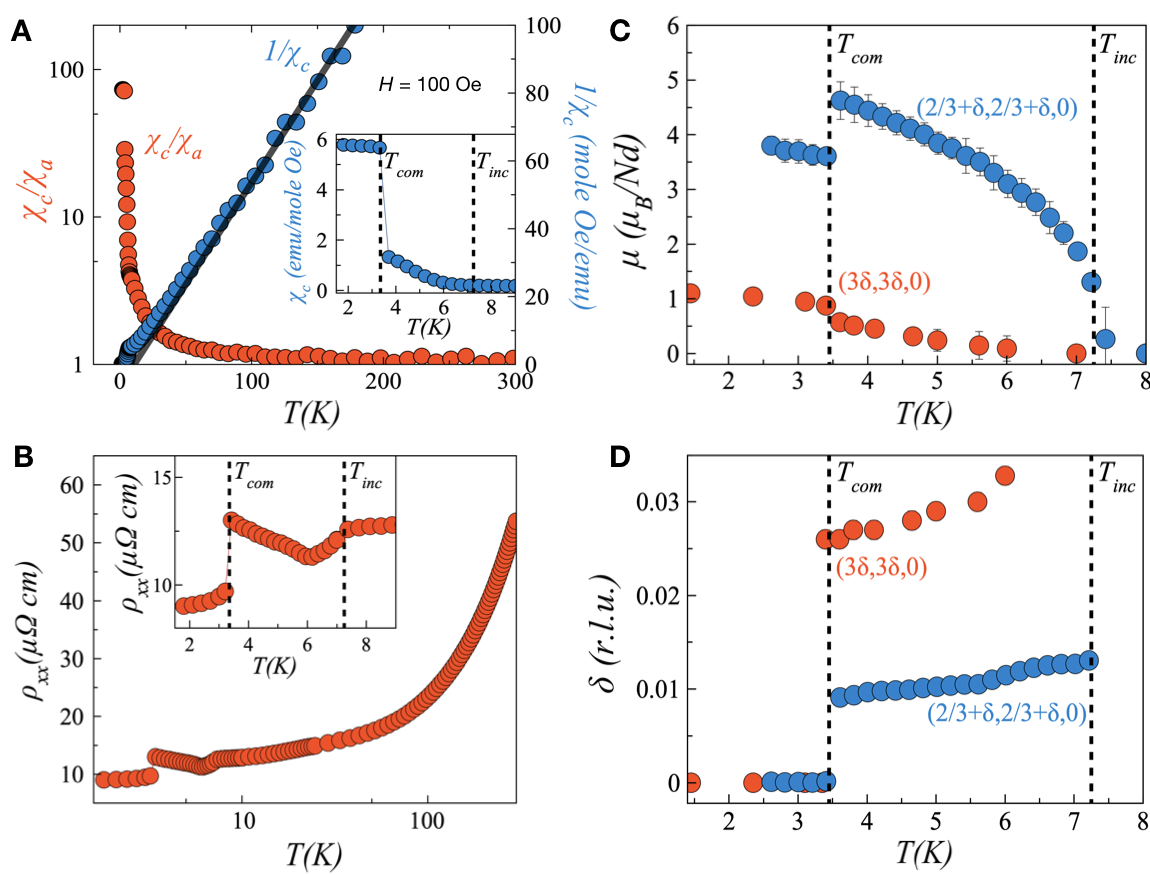}
    \centering
    \caption{\textbf{Magnetic phase transitions in NdAlSi.} (A) temperature dependence of the inverse susceptibility for field applied along the ${\bf c}$ axis ($\chi_c$) as well as its ratio with the in-plane susceptibility ($\chi_a$). (B) is the $ab$ plane longitudinal electrical resistivity of NdAlSi versus temperature. (C) and (D) respectively show the temperature dependence of the ordered moment and the associated wavevectors.}
    \label{fig:Bulk2}
\end{figure}

The temperature dependence of the inverse out-of-plane magnetic susceptibility ($1/\chi_c$) and its ratio with the in-plane susceptibility ($\chi_c / \chi_a$) are reported in Fig.~\ref{fig:Bulk2}A. An isotropic Curie-Weiss fit with $\Theta_{CW}=10.2(8)$~K and $\mu_{eff}=3.65(5)~\mu_B$ describes the data well at high temperature, but fails at low temperature where the system develops an Ising-like easy axis anisotropy with $\chi_c/\chi_a > 80$ at $T=1.8$~K. Comparing Fig.~\ref{fig:Bulk1}D and Fig.~\ref{fig:Bulk2}A shows that the signature of $T_{\mathrm{inc}}$ is clearer in the heat capacity, whereas $T_{\mathrm{com}}$ is most prominent in the magnetic susceptibility data. This is consistent with a putative antiferromagnetic (AFM) transition with $\mathbf{k}\neq\mathbf{0}$ at $T_{\mathrm{inc}}$ and a ferrimagnetic transition with $\mathbf{k}$~=~$\mathbf{0}$ at $T_{\mathrm{com}}$. The magnetic transitions also affect the electrical resistivity (Fig.~\ref{fig:Bulk2}B), where the first small drop at $T_{\mathrm{inc}}$ and the second larger drop at $T_{\mathrm{com}}$ are correlated with the magnetic transitions. As explained below, the resistivity minimum at approximately 6~K correlates with the appearance of an incommensurate order that is precursor to the ferrimagnetic state observed for $T<T_{\mathrm{com}}$.

\section{Neutron Diffraction and Spin Structure}

The magnetic order parameters were determined through single crystal neutron diffraction experiments that are summarized in Fig.~\ref{fig:Bulk2}C and \ref{fig:Bulk2}D. Below $T_{\mathrm{inc}}$, we found antiferromagnetic Bragg peaks that are characterized by an incommensurate $\mathbf{k}_{\rm inc}=(\frac{2}{3}+\delta,\frac{2}{3}+\delta,0)$ ordering wavevector. We also observed magnetic Bragg peaks corresponding to the third harmonics of this wavevector, i.e. $\mathbf{k}=(3\delta,3\delta,0)$ whose intensity smoothly increases upon cooling from $T_{\mathrm{inc}}$ to $T_{\mathrm{com}}$. The incommensurability $\delta(T)$ shows a weak temperature dependence before dropping to zero for $T<T_{\mathrm{com}}$, where a sharp transition to commensurate ferrimagnetic order occurs (Fig.~\ref{fig:Bulk2}D).

\begin{figure}[t]
    \includegraphics[width=470pt]{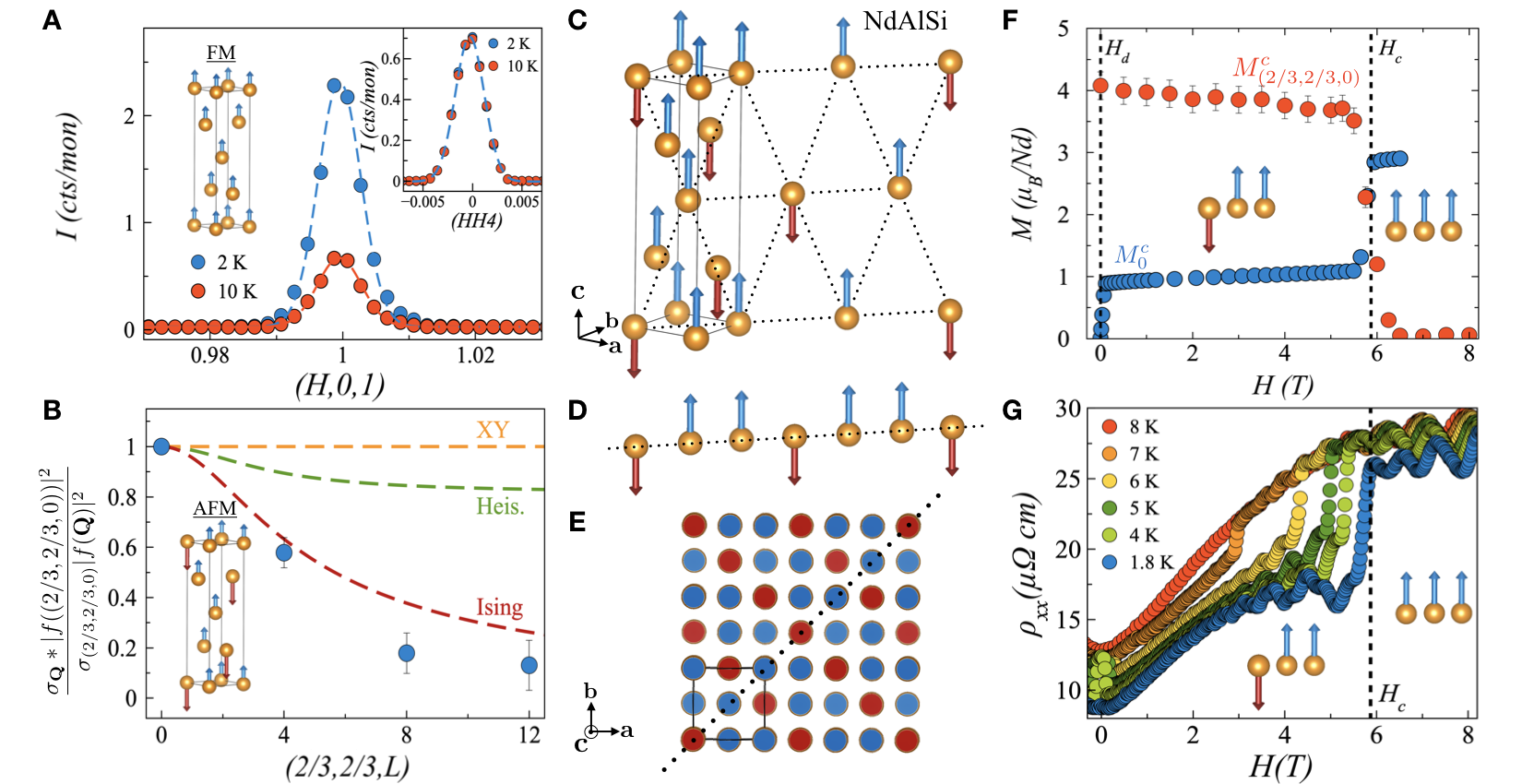}
    \centering
    \caption{\textbf{Commensurate ferrimagnetic spin structure of NdAlSi.} (A) Rocking scans at $\mathbf{Q}=(101)$ and $\mathbf{Q}=(004)$ whose magnetic intensities are consistent with the FM spin structure shown in the left inset. (B) shows the $L$ dependence of the $\mathbf{Q}=(\frac{2}{3}\frac{2}{3}L)$ Bragg peaks at $T=1.6$~K, which have been normalized to the $(\frac{2}{3}\frac{2}{3}0)$ Bragg peaks, and multiplied by the ratio of their form factors. The calculated $L$ dependence for various anisotropies are plotted and reveals the AFM $\mathbf{k}_{\rm com}=(\frac{2}{3}\frac{2}{3}0)$ component is Ising like (see bottom inset). The red spins were refined to 3.8(4)$\mu_B$, the blue spins to half this size. (C) 3D representation of the NdAlSi spin structure, which is obtained by adding the structures shown in the insets of (A) and (B). (D) shows the effective 1D spin configuration of (C) projected to $z$~=~0 and plotted along the [110] direction. (E) shows the spin structure with the $c$ axis coming out of the page. The $c$ axis field dependence of the magnetization as well as the AFM $\mathbf{k}_{\rm com}$ component at $T$~=~1.6~K are shown in (F) and the electrical resistivity in (G) for various temperatures.}
    \label{fig:Neutron}
\end{figure}

The spin polarization of both the FM $\mathbf{k}=(000)$ and the AFM $\mathbf{k}_{\rm com}=(\frac{2}{3}\frac{2}{3}0)$ components of the $T<T_{\rm com}$ spin structure were determined through refinement of magnetic neutron diffraction data (Fig.~S2). The primitive unit cell of NdAlSi contains two Nd ions located at $\mathbf{r_1}=(0,0,0)$ and $\mathbf{r_2}=(\frac{1}{2},0,\frac{1}{4})$ whose spin orientation and size can differ. For the $\mathbf{k}=(000)$ component of the spin structure, however, this difference would produce magnetic Bragg reflections at $\mathbf{k}=(110)$ positions that were not present in our data. Indeed, magnetic Bragg intensity was only detected at $\mathbf{k}=(000)$ positions such as $\mathbf{Q}=(101)$ (Fig.~\ref{fig:Neutron}A). Furthermore, no magnetic Bragg intensity was detected at any of the $\mathbf{Q}=(00L)$ Bragg peaks (such as $\mathbf{Q}=(004)$ shown in the inset of Fig.~\ref{fig:Neutron}A). This precludes ordered in-plane components of the spins. Thus, we conclude that the $\mathbf{k}=(000)$ spin structure component is ferromagnetic with spins oriented along the $\bf c$ direction (see left inset of Fig.~\ref{fig:Neutron}A).

The commensurate $\mathbf{k}_{\rm com}=(\frac{2}{3}\frac{2}{3}0)$ magnetism can be resolved in independent components polarized along the $[110], [1\bar{1}0]$ and $[001]$ real space directions. Each component of the magnetic moment on the $\mathbf{r_1}$ and $\mathbf{r_2}$ Nd sites of the chemical unit cell can either be parallel $(+)$ or antiparallel $(-)$. 
The latter two scenarios can be distinguished through neutron diffraction as $+$ yields ${\bf Q_+}={\bf G} + (\pm1/3,\pm1/3,0)$ Bragg peaks while $-$ leads to ${\bf Q_-}={\bf G} + (\pm2/3,\pm2/3,0)$ peaks where ${\bf G}$ is any allowed nuclear Bragg peak. We found magnetic Bragg scattering at ${\bf Q_-}$ positions that is two orders of magnitude greater than at ${\bf Q_+}$, which clearly points to the $-$ state (Fig.~S2). The Ising anisotropy for this order was established by examining the ${\bf \hat{Q}_-}\cdot\hat{\bf c}$ dependence of the Bragg intensity, which easily distinguishes between isotropic, planar (XY), and Ising like order parameters (Fig.~\ref{fig:Neutron}B). The presence of weak ${\bf Q_+}$ Bragg peaks suggests an in-plane moment of 0.14(5)$\mu_b$/Nd, but further work is needed to properly characterize this component. 

The $\mathbf{k}_{\rm FM}=(000)$ and $\mathbf{k}_{\rm com}=(\frac{2}{3}\frac{2}{3}0)$ components of the NdAlSi spin structure are respectively shown in the inset of Fig.~\ref{fig:Neutron}A and B. A refined moment of 1.1(2)~$\mu_B$/Nd and 3.8(4)~$\mu_B$/Nd  were determined for the FM and AFM components respectively. While the phase relationship between these components is unconstrained by diffraction, a constant moment requirement fixes the phase. The corresponding structure (Fig.~\ref{fig:Neutron}C) can be described as a 1D down-up-up (d-u-u) chain in Fig.~\ref{fig:Neutron}D, where each arrow represents a plane of spins perpendicular to ${\bf k}_{\rm com}$ (Fig.~\ref{fig:Neutron}E).

The magnetization curve for ${\bf H}\parallel {\bf c}$ features two plateaus (Fig.~\ref{fig:Neutron}F). The first plateau for applied field $\mu_0 H$ in the range $\mu_0H_d = 0.1$~T $ < \mu_0 H < \mu_0 H_c = $~5.8~T, with a net magnetization of 1.0(1)~$\mu_B$/Nd is consistent with a single domain d-u-u state. In agreement with the vanishing of $\mathbf{k}_{\rm com}$ Bragg diffraction at $H_c$ (Fig. 3F), the second plateau for $\mu_0 H >\mu_0 H_C=5.8$~T corresponds to the saturated state (u-u-u) with a magnetization of 2.9(1)~$\mu_B$/Nd, which matches the sublattice magnetization of the zero field d-u-u state as determined by diffraction.

Locally, the incommensurate spin structure of NdAlSi is also an Ising d-u-u structure, though with a $\lambda_m=145(8)$~nm long-wavelength amplitude modulation at $T$~=~3.5~K. Incommensurate to commensurate phase transitions occur in many rare earth and actinide intermetallics including UNi$_2$Si$_2$~\cite{Lin1991}, CeNiAs~\cite{Wu2019}, and CeSb~\cite{Rossat1977}. Considering the exchange interactions and the strong easy axis anisotropy, mean field theory shows that amplitude modulated incommensurate spin density wave ordering is generally unstable towards a lower $T$ transition into a commensurate spin structure~\cite{Gignoux1993}. NdAlSi can be understood within this theoretical framework: An amplitude modulated spin density wave with $\mathbf{k}_{\rm inc}=(\frac{2}{3}+\delta,\frac{2}{3}+\delta,0)$ first appears below $T_{\mathrm{inc}}$, its amplitude growing upon cooling until $T_{\mathrm{com}}$. As the amplitude grows, so does the third harmonic $\mathbf{k}=(3\delta,3\delta,0)$, which indicates that the sine wave is "squaring up" and the magnitude of the moment on each site is becoming more constant through the lattice. For an Ising-like incommensurate state, full magnetization on every site is expected at $T=0$ and it requires abrupt soliton-like defects. At $T_{\mathrm{com}}$ the commensurate d-u-u state with constant moment size per site becomes energetically favorable. Though its periodicity deviates from that favored by fermi-surface nesting, the commensurate square wave structure allows all Nd sites to achieve the saturation magnetization (Fig.~\ref{fig:Bulk2}C). 

\begin{figure}[t]
    \includegraphics[width=460pt]{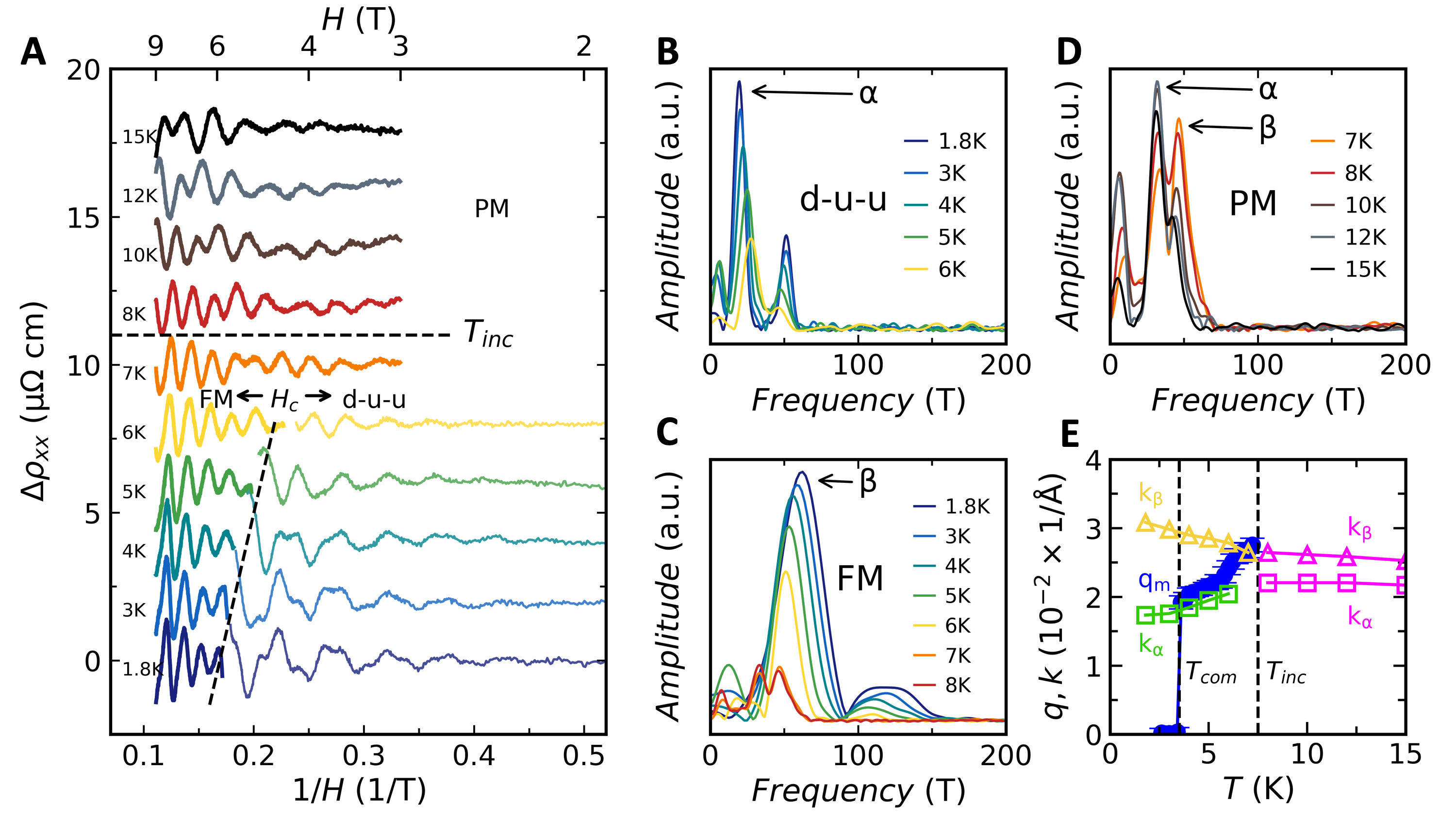}
    \centering
    \caption{\textbf{Quantum oscillations in NdAlSi.} (A) Quantum oscillations (QO) recorded above and below the transition temperature $T_{\mathrm{inc}}(H=0)$ (separated by the horizontal dashed line), above and below the metamagnetic (MM) transition field $H_c$ (separated by the slanted line). The FFT spectra of the oscillations for the PM phase, the AFM $(\frac{2}{3}\frac{2}{3}0)$ phase, and the FM phase are respectively shown in B, C and D. (E) The temperature evolution of the magnetic incommensurability $q_m$ (see Fig.~\ref{fig:Bulk2}), and quantum oscillation frequencies $k_\alpha$ and $k_\beta$ in different states. Note that the frequencies are shown in $\angstrom^{-1}$.}
    \label{fig:QO}
\end{figure}

\section{Quantum Oscillations, Band Structure Calculations, and Fermi Surface}

Magnetic interactions between rare earth ions in metals are mediated by conduction electrons through the Ruderman–Kittel–Kasuya–Yosida (RKKY) mechanism. The often oscillatory character of the ordered state reflects the Fermi surface and the associated Lindhard susceptibility~\cite{Taylor1971,Jensen1991}. To make the connection between the modulated magnetism of NdAlSi and itinerant electrons, we combine DFT calculations and quantum oscillation experiments.

We present the field dependence of resistivity with ${\bf H}\parallel$[001] in Fig.~\ref{fig:Neutron}G. Shubnikov-de Haas (SdH) oscillations are observed at small fields. The pattern of SdH oscillations is visibly different in the ferrimagnetic d-u-u phase for $H<H_c$ compared to the FM u-u-u phase for $H>H_c$. To extract the purely oscillatory part of the resistivity, $\Delta\rho_{xx}$, we subtract a smooth background from the $\rho_{xx}(H)$ curve at each temperature. The resulting $\Delta\rho_{xx}(H)$ curves are plotted as a function of 1/$H$ in Fig.~\ref{fig:QO}A (see also Fig.~S5). Three phases with different oscillatory patterns are separated by dashed lines in Fig.~\ref{fig:QO}A: The paramagnetic (PM) phase above $T_{\mathrm{inc}}$, the ferrimagnetic d-u-u phase below $T_{\mathrm{inc}}$ and $H_c$, and the FM phase below $T_{\mathrm{inc}}$ and above $H_c$. To extract the frequencies of quantum oscillations, we performed a Fourier analysis in each phase, the result of which is shown in Fig.~\ref{fig:QO}B, C and D. The peaks in Fig.~\ref{fig:QO}D reveal the oscillation frequencies in the PM phase with two characteristic frequencies $\alpha$ and $\beta$. Below $T_{\mathrm{com}}$, however, only the $\alpha$ peak survives in the low field d-u-u phase, while only the $\beta$ peak survives in the FM phase (Fig.~\ref{fig:QO}B,C).

We shall now establish a connection between the conduction electrons participating in quantum oscillations and the incommensurate $f$-moment magnetism by comparing the characteristic wavevector inferred from SdH frequencies to the incommensurate ordering wavevector. The SdH frequencies are associated with the conduction electrons and the ordering wavevector with the localized $f$-moment. The SdH oscillation frequencies $F$ are related to the extremal orbit area ($A_{ext}$) on the Fermi surface via the Onsager relation $F=\frac{\phi_0}{2\pi^2} A_{ext}$. From here, we determine the characteristic reciprocal space dimension of the $\alpha$ and $\beta$ orbits using $k_{\alpha,\beta}=\sqrt{A_{ext}}/\pi=\sqrt{2\pi F_{\alpha,\beta}/\phi_0}$ and plot them as a function of temperature along with the incommensurate component of the magnetic wavevector $q_{m}=\delta \sqrt{2} a^*$ in Fig.~\ref{fig:QO}E. All wavevectors vary smoothly with temperature and the similarity of their rate of change with $T$ (specifically  $|dk_{\alpha,\beta}/dT|\approx |dq_{m}/dT|$) is consistent with a link between the incommensurate order and the Fermi surface of NdAlSi. 

\begin{figure*}
\includegraphics[width=1\textwidth]{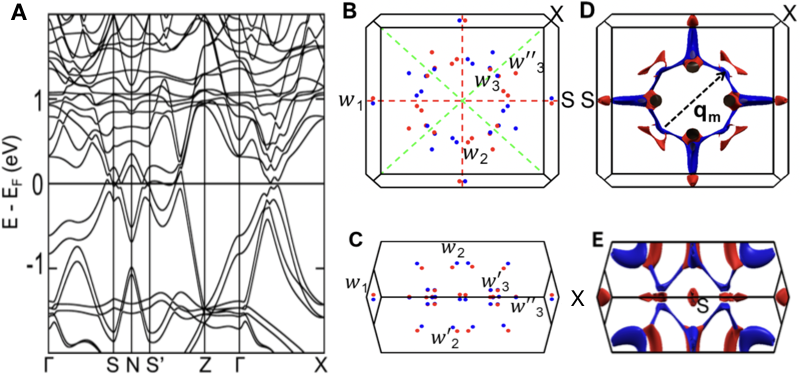}
\centering
\caption{
\textbf{Electronic band structure for the ferromagnetic phase of NdAlSi.} (A) Ferromagnetic PBE+$U$+SOC band structure. (B-C) Ferromagnetic PBE+SOC Weyl nodes in top and side views respectively. The six types of Weyl nodes are marked as W$_1$, W$_2$, W'$_2$, W$_3$, W'$_3$  and W''$_3$. (D) and (E) are respectively the top and side view of the ferromagnetic Fermi surface whose Fermi energy was determined by the quantum oscillations analysis. The red color represents the valence band whereas the blue color is the conduction band.}
\label{dft_band}
\end{figure*}

To explore this possibility, we performed DFT calculations in the various magnetic phases of NdAlSi. For a robust determination of $E_F$ using quantum oscillations data, the DFT results are compared to SdH oscillations in the high field FM state, the band structure of which is shown in Fig.~\ref{dft_band}A. For this calculation, the spin-orbit coupling (SOC) is taken into account and the partially filled $f$ orbitals are included in the valence. A large density of $f$ states is obtained at the Fermi level without onsite Hubbard $U$, but these are pushed away to lower and higher energies in the PBE+$U$+SOC calculations, where we have adopted $U=6$~eV~\cite{Anisimov1997,Yang2020}. In this case, a magnetic moment of $2.94~\mu_{B}/$Nd is obtained, which is indistinguishable from the FM saturation magnetization (Fig. 3F) and the 2.9(1)~$\mu_B$/Nd sublattice magnetization of the commensurate low $T$ state. 

We used the Wannier90~\cite{Pizzi2020} package to accurately interpolate the first-principles band structure (see Methods). From here, we find a total of 56 Weyl nodes in this FM phase whose locations within the first BZ are shown in Fig.~\ref{dft_band}B,C. The colors of the Weyl nodes reflect their chiralities, with red and blue dots representing nodes with chiralities $+1$ and $-1$, respectively. Because the magnetic point group has eight symmetry elements, the Weyl points come in groups of eight nodes that are all degenerate in energy, four with chirality +1 and four with chirality $-1$. We label the nodes as W$_1$ (one group), W$_2$ (two groups), W$_3$ (two groups), W'$_3$ (one group), and W''$_3$ (one group), accounting for a total of 56 Weyl nodes. The type and momentum space location of these Weyl nodes are strongly impacted by the $f$-electron magnetism. For example, Weyl nodes in the FM phase appear close to the glide mirror plane lying along the $\Gamma$--$X$ direction. These arise due to the crossing of spin-up and spin-down bands so they are not stabilized in the PM phase (Fig.S3). Furthermore, Weyl nodes along the $S^\prime$--$Z$ direction change from type I to type II in going from the PM to the FM phase.

Having presented the quantum oscillations and DFT calculations for NdAlSi, we now combine these to obtain the Fermi surface in the FM state (Fig.~\ref{dft_band}D and E). The Fermi level was determined by comparing the calculated $\beta$ frequency from DFT and the experimental value of $k_\beta$ from quantum oscillations, as well as two other quantum oscillation frequencies that were resolved in our high-field experiment (Fig.~S5). The best agreement was obtained with $E_F$-$E_F^{DFT}$~=~+30(3)~meV where $E_F^{DFT}~=~6.7473~eV$ is the charge-neutrality point.

\section{Incommensurate Magnetism and Nesting Wavevector}

The $\mathbf{k}_{\rm inc}=(\frac{2}{3}+\delta,\frac{2}{3}+\delta,0)$ magnetic order of NdAlSi is unique among the reported RAlSi/RAlGe compounds, where so far only commensurate $\mathbf{k}=\mathbf{0}$ orders were reported in zero field~\cite{Puphal2020,Suzuki2019,Yang2020}. To understand the origin of this modulated magnetic order, we note that the Fermi surface for FM NdAlSi includes pockets near ${\bf Q}=(\pm\frac{1}{3},\pm\frac{1}{3},L)$ with both $L$~=~0 and $L\neq0$ (Fig.~\ref{dft_band}D,E). Nesting between symmetrically equivalent diagonal pairs of these pockets might be responsible for the instability of the FM state towards formation of the  $\mathbf{k}_{\rm inc}=(\frac{2}{3}+\delta,\frac{2}{3}+\delta,0)$ spin density wave for $H<H_c$ (see $\mathbf{q_{m}}$ in Fig.~\ref{dft_band}D).

\begin{figure}
\centering
\includegraphics[width=0.9\textwidth]{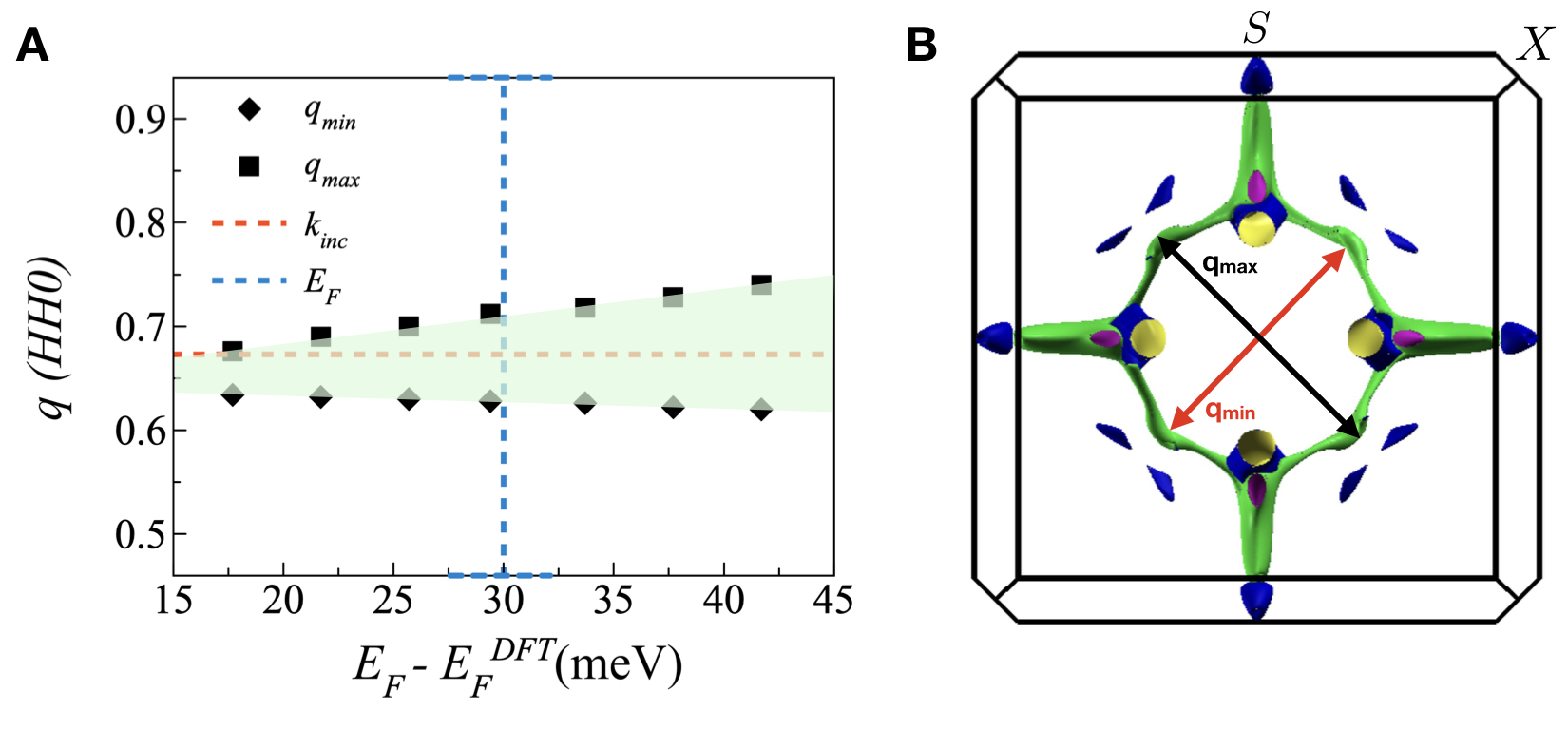}
\caption{\label{Nesting}
\textbf{Nesting vector and the $(\frac{2}{3}+\delta,\frac{2}{3}+\delta,0)$ magnetic order in NdAlSi.}
(A) The black dots represent the minimum ($q_{min}$) and maximum ($q_{max}$) momentum vectors connecting two Fermi pockets of NdAlSi that lie along the (HH0) or $\Gamma-X$ direction of the first Brillouin zone. The green region encompasses all possible nesting vectors as function of $E_F$-$E_F^{DFT}$ where $E_F^{DFT}$ is the charge-neutrality level. The red line indicates the $(\frac{2}{3}+\delta,\frac{2}{3}+\delta,0)$ magnetic wavevector found in NdAlSi and the blue line is the value of $E_F-E_F^{DFT}$ obtained from our quantum oscillation analysis. The horizontal blue bars represent the error bars on $E_F-E_F^{DFT}$. (B) Fermi surface of  FM NdAlSi for $E_F-E_F^{DFT}=33$~meV with arrows indicating ${\bf q}_{min}$ and ${\bf q}_{max}$ used for the analysis presented in A.
}
\end{figure}

A quantitative comparison of the nesting vector $\mathbf{q_{m}}$ and the magnetic ordering vector is presented in Fig.~\ref{Nesting}A. For this figure, we generated the Fermi surface of FM NdAlSi for a wide range of $E_F$ and extracted the extremum positions of the $L$-integrated Fermi pockets lying along the (110) direction. The minimum and maximum extremum positions of these pockets were then multiplied by 2 to define the range of possible nesting vectors (see Fig.~\ref{Nesting}B) bounded by the green region between $\mathbf{q_{min}}(E_F)$ and $\mathbf{q_{max}}(E_F)$ in Fig.~\ref{Nesting}A. The vertical blue bar in Fig.~\ref{Nesting}A represents the value of $E_F$ which was independently determined by the quantum oscillations analysis, while the horizontal red bar is $\mathbf{k}_{\rm inc}$ from neutron diffraction. The red and blue lines meet in the green region where nesting conditions are satisfied. This indicates that the nesting between Fermi pockets lying along the (110) direction could be responsible for the magnetic order observed in NdAlSi. We note that this result is not the fruit of a fine tuning of $E_F$ since the nesting condition is satisfied over a broad range of $E_F$, which extends well beyond the constraints of our quantum oscillations analysis. 

We can then rationalize the low-temperature magnetic order starting from the FM state where the Zeeman term completely dominates and all the spins point along the $c$ axis. Upon decreasing the field to $H_c$, the magnetic susceptibility develops a peak at $\mathbf{k}_{\rm inc}$ due to RKKY interactions that favor the nesting wavevector. At still lower fields the corresponding modulated magnetic order emerges to gap the soft mode. This phenomenology is well established for various rare-earth metallic and intermetallic systems~\cite{Taylor1971,Jensen1991}. 

\section{Discussion and Conclusion} 

NdAlSi is special in that the nesting Fermi pockets contain W$_3$ Weyl nodes 41~meV above E$_F$ (Fig.~\ref{dft_band}B,C). Weyl-mediated RKKY interactions have been studied using perturbation theory~\cite{Hosseini2015,Araki2016,Wang2017,Nikolic2020}. Every pair of Weyl nodes separated by a wavevector ${\bf q}$ contributes Heisenberg, Kitaev, and Dzyaloshinskii-Moriya interactions among local moments. The Heisenberg coupling $J(\delta{\bf r})$ between two moments a distance $\delta{\bf r}$ apart features sign-changing spatial modulations with wavevector ${\bf q}$ and a rapid lost of strength at larger distances $|\delta{\bf r}|>\hbar/\Lambda$, where $\Lambda$ is the momentum cut-off in the relativistic dispersion of Weyl electrons. The overall $J(\delta{\bf r}\to0)<0$ from equal-chirality Weyl nodes promotes a spin-density-wave at ${\bf q}$ while every single node yields a ${\bf q}=0$ ferromagnetic channel through intra-node electron scattering. This latter interaction naturally explains the ${\bf k}_{\rm FM}={\bf 0}$ spin component that ultimately develops in NdAlSi, as it is favored by every Weyl hole and electron pocket.
This suggest that the AFM spin component must come from inter-node scattering processes. The multitude of Weyl nodes in NdAlSi creates many potential AFM ordering channels from which the $\mathbf{k}_{\rm inc} = \left(\frac{2}{3}+\delta, \frac{2}{3}+\delta, 0\right)$ order would arise. To understand this, we note that nearly commensurate nesting wavevectors produce orders that support saturated sublattice magnetization and therefore are energetically favored. Moreover, $\mathbf{k}_{\rm inc}$ approximately connects not one but four symmetry-related pairs of W$_3$ Weyl nodes. Furthermore, the third harmonic 3$\mathbf{k}_{\rm inc}$, which must be present in a sublattice-saturating square wave state and was observed at $(3\delta,3\delta,0)$ (Fig.~\ref{fig:Bulk2}C), is near the $\Gamma$-point. Thus, in contrast to other Weyl-Weyl ordering channels, the third harmonic of $\mathbf{k}_{\rm inc}$ is favored by $q \rightarrow 0$ intra-node scattering. Taken together, these observations provide a coherent provisional explanation of the observed magnetic order.

In conclusion, we have discovered uniaxial incommensurate amplitude modulated magnetic ordering in NdAlSi – an inversion symmetry breaking semimetal with topologically protected Weyl nodes. Upon cooling, the magnetic wavevector ${\bf k}_{\rm inc}=(\frac{2}{3}+\delta,\frac{2}{3}+\delta,0)$ approaches the commensurate point ($\delta\rightarrow 0$) while the third harmonic grows and the incommensurate structure acquires square wave character. This culminates in a first order phase transition to low $T$ commensurate ferrimagnetism with characteristic wavevectors ${\bf k}={\bf 0}$ and ${\bf k}_{\rm com}=(\frac{2}{3}\frac{2}{3}0)$. Low field temperature dependent quantum oscillations of the magnetoresistance links $\delta(T)$ to characteristic linear dimension of the Fermi surface. Through DFT+U calculations we establish the Fermi-surface of the field driven ferromagnetic state – determining the chemical potential through comparison of extremal orbital areas to Subnikov de Haas oscillation frequencies. ${\bf k}_{\rm inc}$ is found to approximately connect four pairs of Weyl nodes and the small associated Fermi pockets. This indicates the magnetic order in NdAlSi is driven by Weyl exchange interactions; inter- and intra-Weyl point processes favoring ${\bf k}_{\rm inc}$ and ${\bf k}={\bf 0}$ respectively. Our work provides a concrete example of collective magnetism driven by Weyl fermions.

\section{Acknowledgments}
This work was supported as part of the Institute for Quantum Matter, an Energy Frontier Research Center funded by the U.S.\ Department of Energy, Office of Science, Basic Energy Sciences under Award No.~DE-SC0019331. The work at Boston College was funded by the National Science Foundation under the award No.~DMR-1708929. A portion of this work was performed at the National High Magnetic Field Laboratory, which is supported by the National Science Foundation Cooperative Agreement No.~DMR-1644779 and the State of Florida. We also acknowledge the support of the National Institute of Standards and Technology, U.S. Department of Commerce. The identification of any commercial product or trade name does not imply endorsement or recommendation by the National Institute of Standards and Technology. Access to MACS was provided by the Center for High Resolution Neutron Scattering, a partnership between the National Institute of Standards and Technology and the National Science Foundation under Agreement No.~DMR-1508249. A portion of this research used resources at the Spallation Neutron Source, a DOE Office of Science User Facility operated by the Oak Ridge National Laboratory. SB thanks Jinwoon Kim for fruitful discussion on the symmetric Wannier functions generations from Wannier90. We thank Yi Li for useful discussions. We are also grateful to Youzhe Chen, Yi Luo, Chris Lygouras, and Yegor Vekhov for their help during neutron scattering experiments.

\section{Methods}
\subsection{Crystal growth.}
We used a self-flux method to grow single crystals of NdAlSi. The starting materials are elemental Nd, Al and Si chunks with composition $\text{Nd}:\text{Al}:\text{Si}=1:10:1$, mixed in an alumina crucible. The crucible was placed in a quartz tube, in which quartz wool was inserted as the strainer, and the tube was sealed under vacuum. The tube was placed in a box furnace, heated to 1000$^\circ$C at 3$^\circ$C/min, held for 12 hours, cooled to 700$^\circ$C at 0.1$^\circ$/min, maintained at 700$^\circ$C for 12 hours, and then centrifuged to remove the excess of Al flux.

\subsection{Transport and Magnetization Measurements.}
Transport properties of NdAlSi were measured in a Quantum Design physical property measurement system (PPMS) Dynacool. Electrical resistivity was measured with a standard four-probe technique, and the heat capacity was measured with a relaxation time method. The dc magnetization was measured using a vibrating sample magnetometer in a Quantum Design MPMS3. We measured quantum oscillations of resistivity in a 35~T DC Bitter magnet inside a $^3$He fridge at a base temperature of 300~mK at the National High Magnetic Field Laboratory in Tallahassee. All samples used for transport measurements were carefully sanded to remove the residual Al-flux and to have the ideal bar geometry for the determination of resistivity.

\subsection{Second harmonic generation}
The second harmonic generation data in the inset of Fig.~\ref{fig:Bulk1}C were taken at normal incidence on the [101] face of as-grown crystals for incoming/outgoing wavelengths of 1500 nm/750 nm as a function of the incoming field polarization and measured for emitted light polarized parallel to [010] crystalline axis. In this geometry, all bulk contributions to SHG from space group 141 are forbidden.

\subsection{Neutron diffraction.}
The structure of NdAlSi was characterized by a single-crystal neutron time-of-flight diffraction experiment on TOPAZ at Oak Ridge National Laboratories. A 1 mm$^3$ sample of NdAlSi was mounted on a goniometer and cooled to 100~K. A 3D neutron diffraction map was acquired for 13 different sample orientations allowing coverage of 1654 Bragg peaks. The nuclear structure factors were extracted from the 3D data sets following a method reported by Schultz \textit{et al.}~\cite{Schultz2014}. The structural refinements were performed within spacegroups 109 and 141 using GSAS-II~\cite{GSAS}. 

The magnetic structure of NdAlSi was determined through various triple-axis neutron diffraction experiments using the BT-7, MACS, and SPINS spectrometers, which are all located at the National Institute of Standard Technology (NIST). These experiments were performed on single crystals of NdAlSi with mass of $\sim$50~mg aligned to collect diffraction data within the $(HK0)$, $(HHL)$ and $(H0L)$ scattering planes for temperatures ranging from 1.5~K to 15~K. A 10~T vertical field magnet was used for the MACS experiment, which was performed with the ${\bf c}$ axis vertical such that the field dependence of the $(\frac{2}{3}\frac{2}{3}0)$ Bragg peak intensity could be measured at 1.6~K for a field applied along the ${\bf c}$ axis. As on MACS, the order parameter measurements were acquired on SPINS with incident and scattered neutron energies of 3.7~meV. Cooled Be filters were employed before and after the sample. Rocking scans within the three magnetic phases of NdAlSi were collected using BT-7 to refine its magnetic structure as a function of temperature. We used PG-filtered 35~meV neutrons for this work. The error bars for the diffraction data correspond to $\pm 1$ standard deviation.

\subsection{DFT.}
The DFT calculations were performed in a plane-wave basis as implemented in the Vienna Ab initio Simulation
Package~\cite{Kresse1996,Kresse1999} with the (PBE)~\cite{Perdew1996} generalized-gradient
exchange-correlation functional. When the $f$ electrons were included in the valence, the on-site correlation was taken into account using the PBE+$U$ approximation with a Hubbard $U=6$~eV~\cite{Anisimov1997}. A strong spin-orbit coupling (SOC) for the Nd atom was also included, which appears to be an important ingredient for topological properties in the presence of magnetism. We obtained a Wannier interpolation of the band structure near the Fermi level using Wannier90~\cite{Pizzi2020} and calculated the Weyl node positions using the WannierTools package~\cite{Wu2018}. A $k$-point grid of 4$\times$4$\times$4 was used to discretize the first Brillouin zone. The relaxation of the atomic positions was carried out keeping the lattice parameters fixed at experimental values with the force criterian 10$^{-2}$\,eV/\AA.

\section{Author contributions} The project was conceived by F.T., D.V., and C.B. Sample synthesis was done by H.-H.Y. and F.T., whereas D.H.T. and B.L. performed and analyzed the second harmonic generation experiments. The heat capacity and bulk susceptibility measurements were carried out by H.-H.Y. and analyzed by J.G. Neutron scattering experiments were performed by J.G., C.B., G.X., Y.Z., J.A.R., and C.M.H. The neutron analysis was performed by J.G. Quantum oscillation measurements were conducted and analyzed by H.-H.Y., F.T., and D.E.G. The density functional theory calculations were carried by S.B. and D.V. P.N. interpreted the data in terms of his theory of exchange interactions mediated by Weyl electrons. The first draft of the paper was written by J.G., H.-H.Y, and S.B to which all authors contributed with comments and edits.

\section{Competing Interests} The authors declare that they have no
competing financial interests.

\end{document}


\maketitle

\begin{affiliations}
 \item Department of Physics and Astronomy and Institute for Quantum Matter, The Johns Hopkins University, Baltimore, Maryland 21218, USA
 \item Department of Physics, Boston College, Chestnut Hill, MA 02467, USA
 \item Department of Physics and Astronomy, Rutgers University, Piscataway, New Jersey 08854, USA
 \item Department of Physics, Temple University, Philadelphia, PA 19122, USA
 \item NIST Center for Neutron Research, National Institute of Standards and Technology, Gaithersburg, MD 20899-6102, USA
 \item Department of Materials Science and Engineering, University of Maryland, College Park, Maryland 20742, USA
 \item Neutron Scattering Division, Oak Ridge National Laboratory, Oak Ridge, Tennessee 37831, USA
 \item National High Magnetic Field Laboratory, Tallahassee, FL 32310, USA
 \item Department of Physics and Astronomy, George Mason University, fairfax, VA 22030, USA

\end{affiliations}

\section{Structural Characterization of NdAlSi}

We have characterized the structure of NdAlSi using powder x-ray diffraction, single crystal neutron diffraction, and Energy-Dispersive X-ray spectroscopy (EDX). The results are summarized in Table.~\ref{table:Struc}. Both neutron diffraction and EDX measurements detected Si and Nd vacancies. The noncentrosymmetric $I4_1md$ (\#109) and centrosymmetric $I4_1/amd$ (\#141) space groups are indistinguishable with X-ray diffraction, and their contrast in neutron diffraction is small. We thus used second harmonic generation (SHG) measurements to distinguish between these space group. SHG is sensitive to a center of inversion in the unit cell. SHG is negligible in a centrosymmetric space group. The large SHG signal reported in the main text points to the noncentrosymmetric space group $I4_1md$ (\#109) and is inconsistent with the centrosymmetric spacegroup $I4_1/amd$ (\#141).

The SHG data were fit to functions appropriate to four different experimental configurations: 1) incoming polarization rotating, output polarizer fixed with polarization parallel to the crystalline $[010]$ axis, denoted $I_{H}(\phi)$; 2) incoming polarization rotating, output polarizer fixed with polarization parallel to the $[10\overline{1}]$ axis, denoted $I_{V}(\phi)$; 3) incoming polarization rotating, outgoing polarizer rotated at 0$^\circ$ angle relative to the incoming polarization, denoted $I_{\parallel}(\phi)$; and 4) incoming polarization rotating, outgoing polarizer rotated with polarization axis at 90$^\circ$ angle relative to the incoming polarization, denoted $I_{\perp}(\phi)$.

\begin{figure}
\centering
\includegraphics[width=0.6\textwidth]{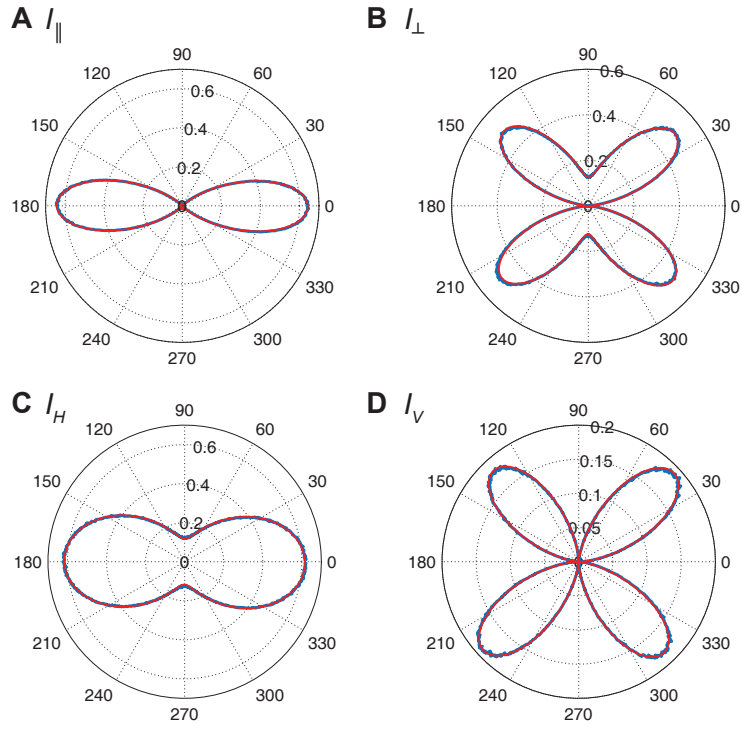}
\caption{\label{SHG}
\textbf{Second-harmonic generation data in NdAlSi.}
The second-harmonic generation (SHG) data for incoming wavelength 1500~nm, outgoing wavelength 750~nm, and fits to bulk electric dipolar SHG in the $C_{4\textrm{v}}$ point group as given by Eqs.~[\ref{eq:eee1}-\ref{eq:eee4}] for (\textbf{A}) $I_\parallel$, (\textbf{B}) $I\perp$, (\textbf{C}) $I_V$, and (\textbf{D}) $I_H$}
\end{figure}

In the electric dipole approximation, the mathematical forms of these various responses for the $[101]$ crystal face in the $I4_1md$ space group ($C_4v$ point group) are given by
\begin{align}
&I^{eee}_{\parallel}(\phi) = \frac{1}{32} \cos ^2(\phi ) \left[ (-2 \chi^{eee}_{xxz}-\chi^{eee}_{zxx}+\chi^{eee}_{zzz})\cos (2 \phi )+6 \chi^{eee}_{xxz}+3 \chi^{eee}_{zxx}+\chi^{eee}_{zzz}\right]^2\label{eq:eee1}\\
&I^{eee}_{\perp}(\phi) = \frac{1}{8} \sin ^2(\phi )\left[ (-2 \chi^{eee}_{xxz}+\chi^{eee}_{zxx}+\chi^{eee}_{zzz}) \cos ^2(\phi )+2 \chi^{eee}_{zxx} \sin ^2(\phi )\right]^2\label{eq:eee2}\\
&I^{eee}_{H}(\phi) = \frac{1}{8} \left[(2 \chi^{eee}_{xxz}+\chi^{eee}_{zxx}+\chi^{eee}_{zzz})\cos ^2(\phi )+2 \chi^{eee}_{zxx} \sin ^2(\phi )\right]^2\label{eq:eee3}\\
&I^{eee}_{V}(\phi) = 2\left[\chi^{eee}_{xxz} \sin(\phi ) \cos(\phi )\right]^2\label{eq:eee4}
\end{align}
The data were fit to expressions~[\ref{eq:eee1}-\ref{eq:eee4}] accounting for a rotation of the sample axes relative to the laboratory x-axis to produce excellent fits to the data, as seen in Fig.~\ref{SHG}. The SHG susceptibilities extracted from those fits are $\chi^{eee}_{xxz}~=~-115\pm3pm/V$, $\chi^{eee}_{zxx}~=~94\pm 2pm/V$, and $\chi^{eee}_{zzz}~=~564\pm5pm/V$. The competing space group assignment $I4_1/amd$ (point group $D_{4h}$) is centrosymmetric and thus should not produce as strong of a SHG response as we have shown here.

\section{Neutron scattering of NdAlSi}
\begin{figure}
\centering
\includegraphics[width=0.9\textwidth]{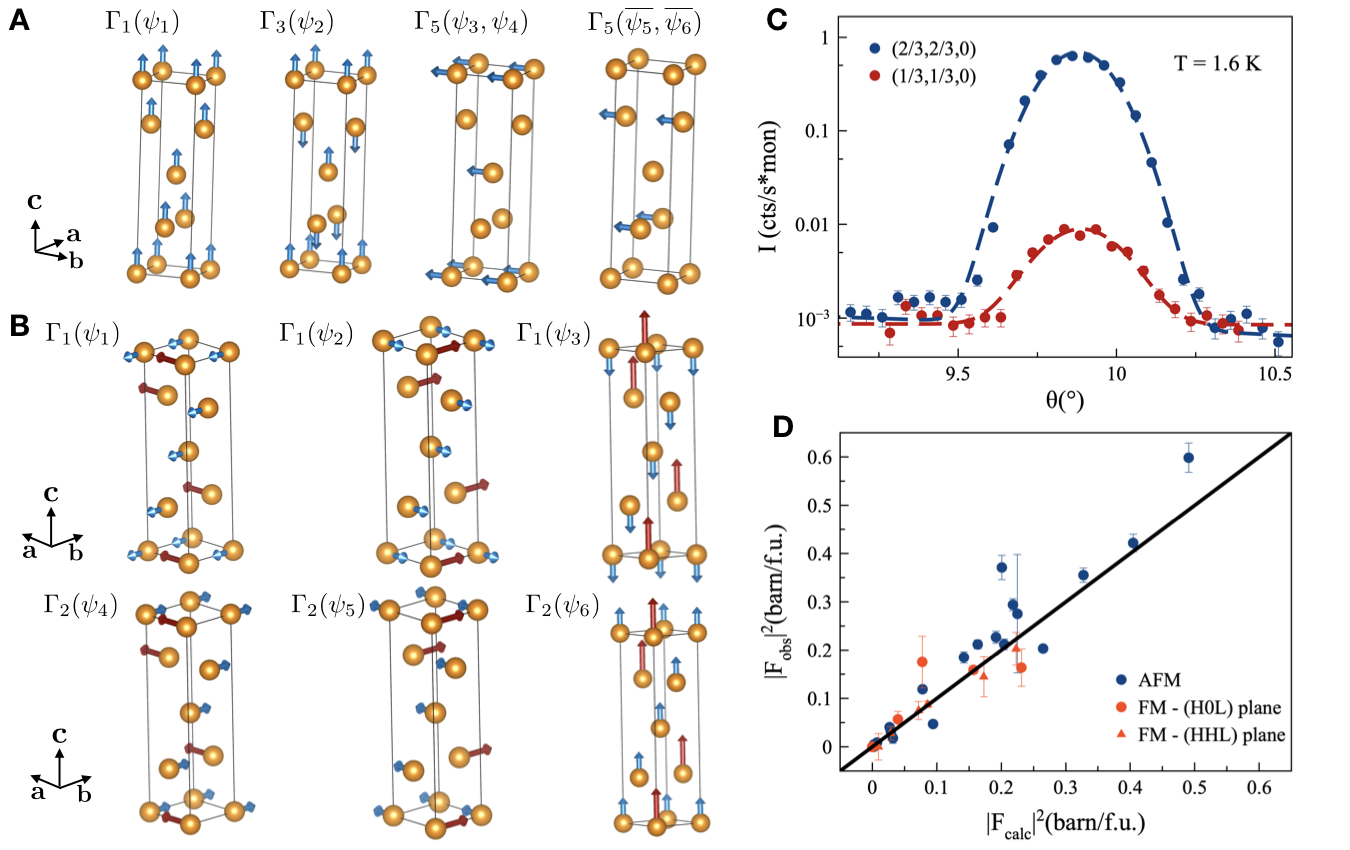}
\caption{\label{NeutronApp}
\textbf{Magnetic structure refinement of NdAlSi.}
Real part of the basis vectors for a $\mathbf{k}=(000)$ and $\mathbf{k}=(\frac{2}{3}\frac{2}{3}0)$ spin structure within space group 109 are shown in A and B respectively. Rocking scans at the $\mathbf{Q}=(\frac{2}{3}\frac{2}{3}0)$ and $\mathbf{Q}=(\frac{1}{3}\frac{1}{3}0)$ Bragg positions are compared in C. Note that $\theta$ for the $\mathbf{Q}=(\frac{1}{3}\frac{1}{3}0)$ Bragg peak was translated by 4.95\degree. The observed magnetic structure factor as a function of the calculated structure factor for both the FM $\mathbf{k}=(000)$ and the AFM $\mathbf{k}=(\frac{2}{3}\frac{2}{3}0)$ components are plotted in D.}
\end{figure}

We refined the spin polarization of the $\mathbf{k}=(000)$ magnetic structure of NdAlSi by acquiring rocking scans at 17 symmetrically non-equivalent $\mathbf{k}=(000)$ Bragg positions covering both the (H,0,L) and (H,H,L) planes. The nuclear and magnetic contributions to the Bragg diffraction were distinguished by collecting rocking scans within both the paramagnetic phase at 10~K and in the commensurate phase at 1.6~K. The Cooper-Nathans formalism was used to calculate the resolution function of our triple-axis experiments~\cite{cooper1967resolution} and convert the integrated intensities of rocking scans to fully integrated Bragg intensities. Symmetry analysis reveals three possible irreducible representations (irreps) to describe the $\mathbf{k}=(000)$ magnetic structure below T$_{c1}$~\cite{Fullprof}: $\Gamma_1$ and $\Gamma_3$ that respectively corresponds to ferromagnetic and antiferromagnetic structures where the spins are oriented along the $c$ axis, and $\Gamma_5$ that corresponds to structures where the spins lie in the $ab$ plane. The real part of the basis vectors associated to each irrep are shown in Fig.~\ref{NeutronApp}A. Generally, the $\mathbf{k}=(000)$ spin structure of NdAlSi can be described as any linear combinations of all the basis vectors within the three irreps. However, as discussed in the main text, $\Gamma_3$ and $\Gamma_5$ respectively produces magnetic Bragg reflections at $\mathbf{Q}=(110)$ and $\mathbf{Q}=(00L)$ positions that were not observed in our diffraction experiments. The final refinement of the neutron diffraction data is plotted in Fig.~\ref{NeutronApp}C and it corresponds to the $\Gamma_1$ structure with $\mu_{FM}=1.1(2)\mu_B$. We note that we also collected 10 rocking scans at positions corresponding to the $\mathbf{k}=(3\delta,3\delta,0)$ component of the incommensurate spin structure of NdAlSi, and we found that the structure factor of this component matches that of the commensurate $\mathbf{k}=(000)$ component. 

We also determined the spin polarization of the AFM components of the spin structure in both the commensurate $(\delta=0)$ and incommensurate phases $(\delta\neq 0)$. To do so, rocking scans at 46 symmetrically non-equivalent Bragg positions were collected within the manifold of Bragg peaks $\mathbf{Q_+}=\mathbf{G} \pm (\frac{1}{3}+\delta,\frac{1}{3}+\delta,0)$ and $\mathbf{Q_-}=\mathbf{G} \pm (\frac{2}{3}+\delta,\frac{2}{3}+\delta,0)$ for both $T=1.6$~K and 6~K. Here $\mathbf{G}$ refers to the manifold of nuclear Bragg peaks. The symmetry analysis of the $\delta=0$ phase reveals six possible basis vectors divided into two different irreps ($\Gamma_1$ and $\Gamma_2$)~\cite{Fullprof}, for which their real parts are shown in Fig.~\ref{NeutronApp}B. The two Nd ions located at $\mathbf{r_1}$=(0,0,0) and $\mathbf{r_2}$=(1/2,0,1/4) in the chemical unit cell have their spins anti-parallel to each other for spin structures described by $\vec{\psi_1}$+$\vec{\psi_2}$, $\vec{\psi_4}$-$\vec{\psi_5}$, or $\vec{\psi_3}$. These anti-parallel spin structures leads to strong $\mathbf{Q_-}$ peaks and no intensity at $\mathbf{Q_+}$ peaks. On the other hand, spin structures described by $\vec{\psi_1}$-$\vec{\psi_2}$, $\vec{\psi_4}$+$\vec{\psi_5}$, or $\vec{\psi_6}$ have Nd spins at $\mathbf{r_1}$ and $\mathbf{r_2}$ that are parallel. This situation leads to strong $\mathbf{Q_+}$ peaks and no intensity at $\mathbf{Q_-}$ peaks. As seen in Fig.~\ref{NeutronApp}C, we observed intensities at $\mathbf{Q_-}$ positions that are two order of magnitude greater than at $\mathbf{Q_+}$ so the spin structure is predominantly of the anti-parallel variety. Fig.~3B of main paper shows it is described by the Ising $\vec{\psi_3}$ basis vector. The final refinement is shown in Fig.~\ref{NeutronApp}D using a $\vec{\psi_3}$ basis vector with a moment of $\mu_{AFM}~=1.9(2)\mu_B$. A 4(3)$\degree$ shift away from a perfect Ising anisotropy is indicated by weak, but finite $\mathbf{Q_+}$ Bragg peaks. However, additional work is needed to properly quantify this possible spin canting. Finally, the relative intensities of the $\mathbf{k}=(\frac{2}{3}+\delta,\frac{2}{3}+\delta,0)$ Bragg peaks within the incommensurate phase are indistinguishable from those of the Ising-like commensurate phase so no spin reorientation was observed above $T_{inc}$. 

For the AFM component, the spatial variation of the Nd moments is expressed as:
\begin{equation}
\mu_{AFM}(\mathbf{r})=1.9(2)(\exp{(i [(\frac{2}{3}\frac{2}{3}0) \cdot \mathbf{r} +  \theta])} + \exp{(-i [(\frac{2}{3}\frac{2}{3}0) \cdot \mathbf{r} + i \theta])}).
\end{equation}
This expression includes both the $\mathbf{k}=(\frac{2}{3}\frac{2}{3}0)$ and $\mathbf{k}=(\bar{\frac{2}{3}}\bar{\frac{2}{3}}0)$ components as required for the magnetic moment to be real for all $\mathbf{r}$. While the diffraction pattern is independent of $\theta$, the real space spin structure does depend on $\theta$. For $\theta~=\pi$ the spin structure can be described as ($0$-up-down) where $0$ means there is no net magnetization on this site, whereas a $\theta~=0$ phase shift leads to an (up-down-down) spin structure. Within the commensurate phase, once the FM component of the structure is added ($\mu_{FM}$~=~1.1(2)$\mu_B$), $\theta~=0(4)\degree$ is the only phase that allows for all the Nd moments to not exceed the 2.9(1)$\mu_B$ saturated moment determined by the magnetization data. The intensities of the $\mathbf{k}=(\frac{2}{3}+\delta,\frac{2}{3}+\delta,0)$ Bragg peaks increase above $T_{inc}$. This requires a phase shift of $\theta~=12(5)\degree$ to respect the same condition within the incommensurate phase.

\section{Electronic band structure of NdAlSi}

\begin{figure}
\centering
\includegraphics[width=\linewidth]{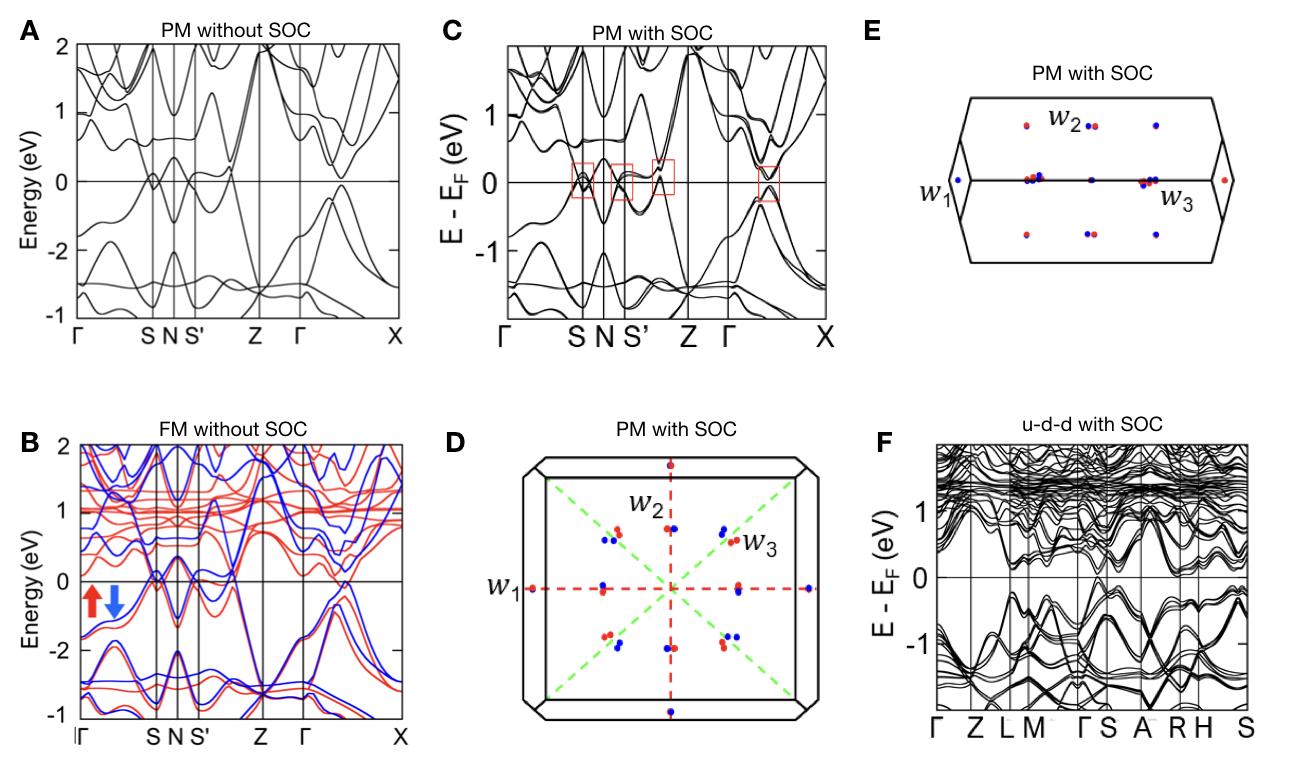}
\caption{\label{dft_band}
\textbf{Band structure of NdAlSi from first-principles calculations.}
Paramagnetic (PM) state in the PBE approximation, with Nd $f$ states
in the core for panel A, and ferromagnetic (FM) state in the PBE+$U$ approximation, with Nd $f$ states
in the valence for panel B. Panel C is the paramagnetic PBE+SOC band structure where the positions of the Weyl nodes in the first Brillouin zone are shown in panel D and E for this calculation. Three types of Weyl nodes W$_1$, W$_2$ and W$_3$ are marked. The mirror planes are shown by dotted lines. Panel F is the PBE+SOC band structure calculation for the u-d-d phase of NdAlSi.}
\end{figure}

In this section, we present a detailed characterization of the band structure of NdAlSi for the different magnetic phases discussed in the main manuscript. We first analyze the band structure without spin-orbit coupling (SOC) for which the nonmagnetic case is shown in Fig.~\ref{dft_band}A. For these calculations, the Nd $f$ states were kept in the core. Linear crossings appear along the high-symmetry lines of the first Brillouin zone. The inclusion of ferromagnetism in the calculation, now including the Nd $f$ orbitals in the valence, induces a spin-exchange splitting between majority and minority spin channels as shown in Fig.~\ref{dft_band}B. Just like the nonmagnetic case, multiple linear crossing points appear along the high-symmetry direction. The majority and minority spin channels are colored in blue and red respectively. The majority and minority spin bands have a simple crossing along the $\Gamma-X$ line, while the pattern of crossings is more complex along $S^{\prime}-Z$, with a tilted crossing very close to the Fermi level, which is in contrast to the nonmagnetic counterpart in Fig.~\ref{dft_band}A.

We now investigate the band structure of NdAlSi including SOC, starting with the nonmagnetic band structure shown in Fig.~\ref{dft_band}C. Because the $f$ orbitals are frozen in the core, they play no role in the active states near the Fermi energy. In this situation, a tiny density of states of $N(E_{F})=0.0012$\,states/eV-cell appears at the Fermi level, composed mainly of Nd $d$, As $s$ and $p$, and Si $s$ and $p$ orbitals. Without SOC in the band structure there are several type-I linear band crossings near the Fermi level along the high-symmetry lines in the BZ (Fig.~\ref{dft_band}A). Due to the presence of SOC in the nonmagnetic calculation, the linear band crossing points highlighted by the red boxes in Fig.~\ref{dft_band}C are gapped out along the high-symmetry directions, but Weyl points now appear slightly off the high-symmetry planes in the BZ. The Wannier90~\cite{Pizzi2020} based tight-binding calculation confirms that there are 40 resulting Weyl nodes in the entire BZ, and the locations of the Weyl nodes are plotted in Figs.~\ref{dft_band}D-E for the top and side views of the BZ, respectively. Four pairs of nodes denoted W$_1$ are located near the zone-boundary $S$ points, with two pairs near the $m_x$ and two near the $m_y$ mirror planes shown by red dashed lines. Four more pairs named W$_2$ lie above and below the $k_z=0$ plane, also near the $m_{x}$ and $m_{y}$ planes, but now in the interior of the BZ, shown by dotted lines. Another twelve pairs of Weyl nodes denoted W$_3$ are arranged around the $m_{xy}$ and $m_{x\bar{y}}$ mirror planes, shown by dashed green lines. The colors of the Weyl nodes reflect their chiralities, with red and blue dots representing nodes with chiralities $+1$ and $-1$ respectively. 

The band structure calculation for the FM phase of NdAlSi are presented in the main manuscript within the PBE+$U$+SOC approximation (Fig.~5). As compared to the nonmagnetic case, the Weyl nodes W$_1$ near the $k_{z}=0$ plane are unaffected by the magnetization along [001], but the nodes lying off the $k_{z}$ plane along the $S^\prime-Z$ direction undergo further splitting around the $m_{x}$ and $m_{y}$ mirror planes, which we defined as two subtypes: W$_2$ nodes close to the $m_x$ and $m_y$ mirror planes, and W'$_2$ nodes away from them. Furthermore, splitting of the W$_3$ nodes along the $m_{xy}$ and $m_{x\bar{y}}$ mirror planes is observed, which gives rise to new pairs of Weyl nodes categorized into three types: W$_3$ nodes close to the $k_z=0$ plane, W'$_3$ off the $k_z=0$ plane, and W''$_3$ close to the $k_z=0$ plane but away from the location of the W$_3$ nodes. To visualize their dispersions, we plot the projected band structure in the vicinity of some representative Weyl nodes in Fig.~\ref{dft_weyl}. Figure~\ref{dft_weyl}A shows the projected band structure plotted along the $k_{y}$ direction with Fermi level at the energy position of Weyl nodes of type W$_1$ with positive chirality, located 98.6~meV above the charge-neutrality Fermi level. The corresponding two-dimensional Fermi surface projected in the $k_{x}-k_{y}$ plane is shown in Fig.~\ref{dft_weyl}B. Similarly, the band structure for a Fermi level aligned with Weyl node W$_2$ is projected along $k_{y}$ in Fig.~\ref{dft_weyl}C, and the same is done for W'$_2$ in Fig.~\ref{dft_weyl}E. The Fermi surfaces projected in the $k_{x}-k_{y}$ plane at these energy positions are shown in Fig.~\ref{dft_weyl}D and F respectively. The cyan-colored circles indicate the Weyl node positions for which the band structures are plotted. The band structure in Fig.~\ref{dft_weyl}E, around the W'$_2$ nodes, appear to be of type II (hole and electron pocket touch each other at the node). The Weyl node W$_2$ in Fig.~\ref{dft_weyl}C is at 9~meV and W'$_2$ in Fig.~\label{dft_weyl}E is at 72~meV above the charge-neutrality Fermi level. Finally, the band structures with Fermi level at the energy positions of the W$_3$, W'$_3$ and W''$_3$ nodes are shown in Fig.~\ref{dft_weyl}G, I and K respectively with the corresponding Fermi surfaces projected in the $k_{x}-k_{y}$ plane shown in Fig.~\ref{dft_weyl}H, J and L. The W$_3$, W'$_3$ and W''$_3$ nodes are at 41 meV, 25 meV, and 48~meV above the charge-neutrality Fermi level, respectively.

Finally, the PBE+$U$+SOC band structure for the commensurate AFM d-u-u phase of NdAlSi is shown in Fig.~\ref{dft_band}F for which the presence of $54$ Weyl nodes is also predicted. The electronic state appears to have a semimetallic character whose details are extremely sensitive to the structural parameters of NdAlSi, which we know from our neutron diffraction experiment is further complicated by the presence of Si and Nd vacancies on the order of few $\%$ (Table.~\ref{table:Struc}). Thus, a robust characterization of the Weyl nodes in the d-u-u phase requires further analysis that is beyond the scope of the present manuscript.

\section{Quantum Oscillations, Determination of $E_F$, and nesting vector in NdAlSi}

In the main text, we have determined the Fermi surface of NdAlSi in its ferromagnetic (FM) phase using the quantum oscillations (QOs) observed in our high-field resistivity measurements. Fig.~\ref{QO}A shows the resistivity of NdAlSi measured up to 35~T, where QOs are apparent at high fields. A smooth background was fitted to the data for fields above the metamagnetic (MM) transition, which is marked by a jump in resistivity. To isolate the QOs, we subtracted a smooth background for each temperature. The resulting resistivity data are shown in Fig.~\ref{QO}B. These data show that the $\Delta R = 0$ axis passes through the center of the oscillations, and a monotonic decrease of the oscillatory amplitudes at all fields. Both observations justify our choice of the smooth background. 

Next, we performed a fast Fourier transformation (FFT) of the data in Fig.~\ref{QO}B. The resulting FFT spectrum is plotted in Fig.~\ref{QO}C. Based on this spectrum, three distinct frequencies $\delta$, $\beta$, and $\gamma$ are identified, together with the second harmonic of the $\beta$ frequency ($2\beta$). Note that the $\beta$ frequency shifts downward as the temperature increases, which agrees with the data presented in the main text (Fig.~4C). We also determined the effective masses for the three Fermi pockets associated with the QOs peak at $\delta$, $\beta$, and $\gamma$. To do this, we fit the standard Lifshitz-Kosevich (LK) formula~\cite{shoenberg_magnetic_2009,willardson1967physics} to their QOs amplitudes as a function of temperature (Fig. \ref{QO}D). The frequencies and effective masses extracted from Fig.~\ref{QO}C and D are listed in Table~\ref{table:SdH}.

With the experimental QOs frequencies analyzed, we now turn to the theoretical QO frequencies given by DFT calculations, and determine the $E_F$ by matching the experiment with the theory. Each $E_F$ corresponds to a specific Fermi surface, which generates a set of QO frequencies. We start from the $E_F$ determined by DFT ($E_F^{DFT}~=~6.7473~eV$, the neutrality point), adjust the $E_F$ below and above $E_F^{DFT}$, compute QO frequencies for different $E_F$'s until the calculated QO frequencies match with those we observed in the high-field experiment. For $E_F$ being 30~meV and 33~meV above $E_F^{DFT}$, the QO frequencies comparable to $\delta$, $\beta$ and $\gamma$ are listed in Table \ref{table:SdH}. As shown in Table \ref{table:SdH}, both the frequencies and effective masses are in decent agreement with the experimental values. We thus choose $E_F=E_F^{DFT}+30(3)$ meV as the appropriate $E_F$ for the FM phase of NdAlSi.

\break

\begin{table}[]
\centering
\begin{tabular}{|c|c|c|c|c|c|c|c|}
\hline
Ions & \hspace{0.5cm}$x$\hspace{0.5cm}  & \hspace{0.5cm}$y$\hspace{0.5cm}    & \hspace{0.5cm}$z$\hspace{0.5cm}         & Occ    & $U_{11}$    & $U_{22}$    & $U_{33}$    \\ \hline \hline
Nd   & 0  & 0   & 0        & 0.96(3) & 0.0039(1) & 0.0028(1) & 0.0019(1) \\ \hline
Al   & 0  & 0   & 0.5841(6) & 1.0(2)  & 0.0070(1) & 0.0019(1) & 0.0002(8) \\ \hline
Si   & 0  & 0   & 0.4182(5) & 0.94(3) & 0.0029(1) & 0.0026(1) & 0.0019(1) \\ \hline
\end{tabular}
\caption{\label{table:Struc} \textbf{Structural parameters of NdAlSi.} The $x$, $y$, and $z$ positions of the Nd, Al and Si ions are tabulated with their occupation number (Occ) and their anisotropic displacement parameters ($U_{11}$, $U_{22}$ and $U_{33}$). These parameters were obtained from structural refinement of the single crystal neutron diffraction data assuming the $I4_1md$ (\#109) space group with $a=b=4.1972(1)~\angstrom$, and $c=14.4915(6)~\angstrom$ refined at $T=100K$.}
\end{table}

\clearpage

\clearpage

\clearpage

\clearpage

\begin{figure}
\centering
\includegraphics[width=\linewidth]{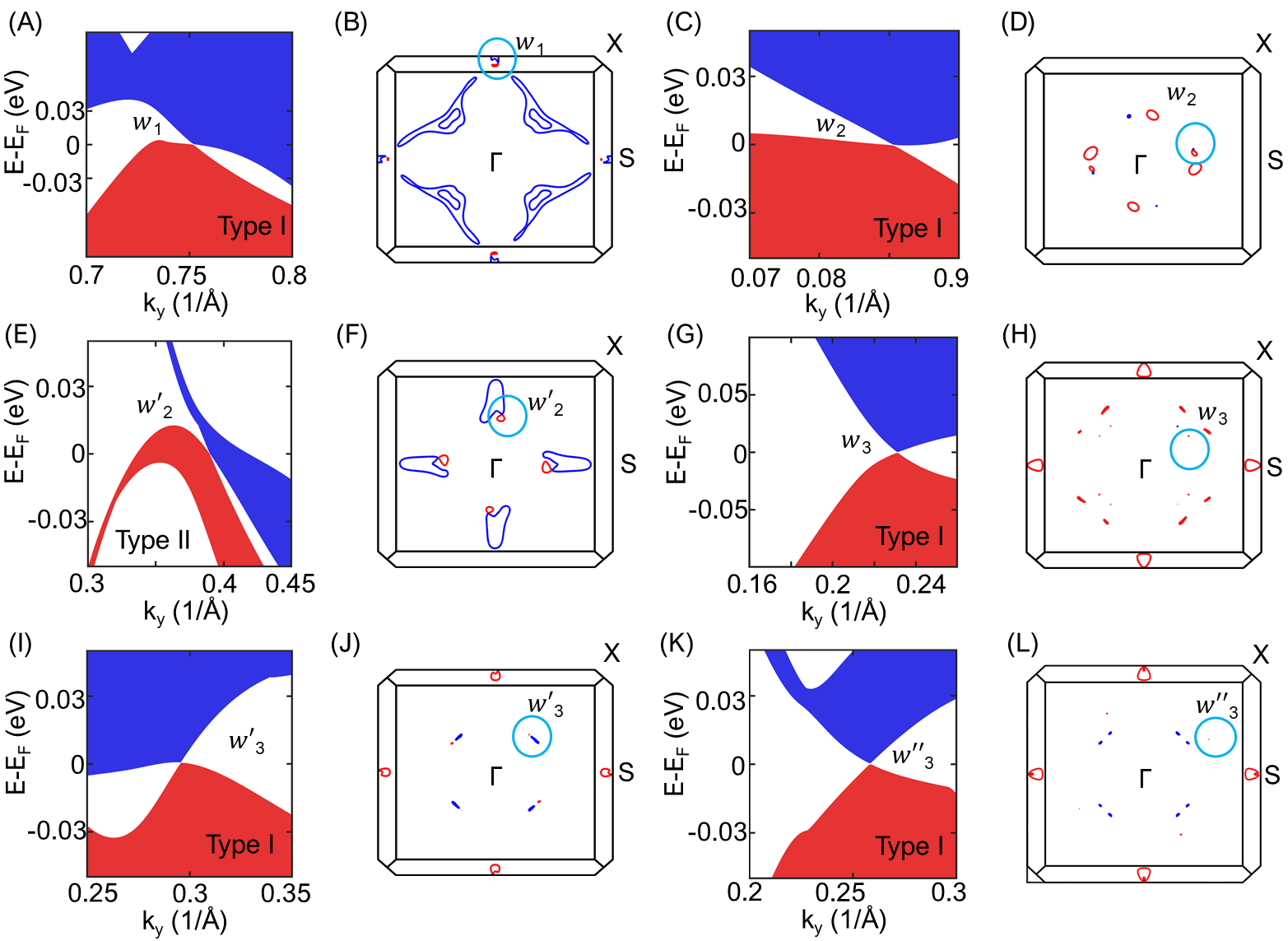}
\caption{\label{dft_weyl}
\textbf{Band structures, Fermi surfaces, and Weyl node positions as
calculated within PBE+$U$+SOC.}
(A-B) Weyl nodes W$_1$,
(C-D) Weyl nodes W$_2$.
(E-F) Weyl nodes W'$_2$.
(G-H) Weyl nodes W$_3$.
(I-J) Weyl nodes W'$_3$.
(K-L) Weyl nodes W''$_3$.}

\end{figure}

\begin{figure}
\centering
\includegraphics[width=0.9\textwidth]{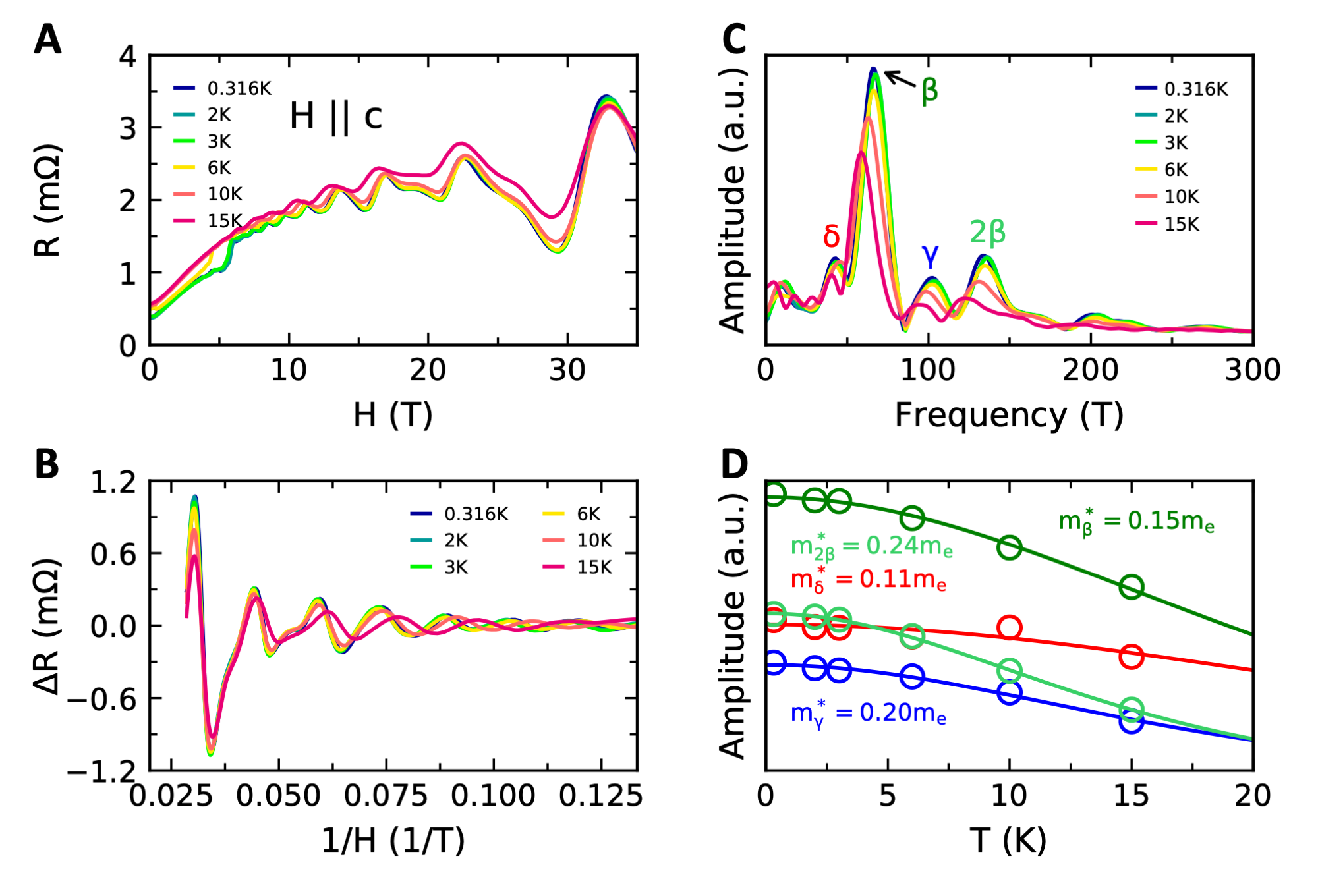}
\caption{\label{QO}
\textbf{High-Field Quantum Oscillations in NdAlSi.}
Resistivity measured up to 35 T at various temperatures with field along the c-axis in A. Shubnikov--de Haas (SdH) oscillations appear and grow as the field increases. SdH oscillations as a function of $1/H$ at different temperatures in B. The change in the oscillatory frequencies can be seen from the shift of peak positions as the temperature increases. FFT spectrum based on the oscillations in panel B is reported in C. Three distinct Fermi pockets are identified: $\delta$, $\beta$, and $\gamma$, and their frequencies at 0.316 K are 40 T, 66 T, and 101 T respectively. The peaks marked by $2\beta$ ($F_{2\beta}=135$ T at 0.316 K) are identified as the second harmonics of $\beta$. Note that $\delta$ frequency is different from $\alpha$ frequency ($F_\alpha = 20$ T at $T=1.8$~K) in the d-u-u phase. The effective masses extracted for each Fermi pocket using a standard Lifshitz-Kosevich formula are reported in D.
}
\end{figure}

\clearpage

\begin{table}[]
\centering
\begin{tabular}{|c|c|c|c|c|c|c|}
\hline
Sources        & $F_\delta$ (T) & $\text{m}_\delta^*$ (m$_e$) & $F_\beta$ (T) & $\text{m}_\beta^*$ (m$_e$) & $F_\gamma$ (T) & $\text{m}_\gamma^*$ (m$_e$) \\ \hline \hline
QO             & 40(5)      & 0.11(2)             & 66(5)     & 0.15(1)            & 101(5)     & 0.20(1)             \\ \hline
DFT, $+30$ meV & 34(1)     & 0.21(1)             & 81(1)    & 0.14(1)            & 103(4)    & 0.20(3)             \\ \hline
DFT, $+33$ meV & 40(1)     & 0.23(1)             & 78(1)    & 0.14(1)            & 98(4)     & 0.20(3)             \\ \hline
\end{tabular}
\caption{\label{table:SdH} \textbf{Comparison of Quantum Oscillation (QO) Frequencies.} The frequencies of the $\delta$, $\beta$, and $\gamma$ pockets derived from resistivity measured up to 35 T, are listed with the results from the DFT calculations. $\delta$ is identified to be an electron pocket, while $\beta$ and $\gamma$ are hole pockets. The ``$+$XX meV'' in the column ``Sources'' means the shift in $E_F$ from the DFT-determined $E_F$ for that particular calculation. As a result of such shift, the frequencies change between the calculations ``DFT, $+30$ meV'' and ``DFT, $+33$ meV''. The error in $E_F$ is determined to be 3 meV based on the change in $E_F$ from matching $F_{\gamma}$ to $F_\delta$.
}
\end{table}